%% file: ase19-final.tex
\documentclass[10pt, conference]{IEEEtran}

\IEEEoverridecommandlockouts
\usepackage{booktabs}
\usepackage{balance}
\usepackage[utf8]{inputenc}
\usepackage{amsmath}
\usepackage{amsfonts}
\usepackage{amssymb}
\usepackage{amsthm}
\usepackage{graphicx}
\usepackage{listings}
\usepackage{algorithm}
\usepackage[noend]{algpseudocode}
\usepackage[utf8]{inputenc}
\usepackage[english]{babel}
\usepackage{xspace}
\usepackage{tabularx}
\usepackage{multirow}
\usepackage{mathrsfs}
\usepackage[table,xcdraw]{xcolor}
\usepackage{xcolor}
\usepackage{listings}
\usepackage{paralist}
\usepackage{subcaption}
\usepackage{graphicx}
\usepackage[skins]{tcolorbox}
\let\labelindent\relax
\usepackage{enumitem,kantlipsum}
\usepackage{mathtools}

\DeclarePairedDelimiter\floor{\lfloor}{\rfloor}
\newtcolorbox{myframe}[2][]{%
  enhanced,colback=white,colframe=black,coltitle=black,
  sharp corners,
  toprule=1.0pt,
  rightrule=0.3pt,
  leftrule=0pt,
  bottomrule=0pt,
  fonttitle=\itshape\scshape\large,
  left=0pt,right=5pt,top=5pt,bottom=3pt,
  attach boxed title to top right={yshift=-0.3\baselineskip-0.4pt,xshift=-5mm},
  boxed title style={tile,size=minimal,left=0.2mm,right=0.5mm,
    colback=white,before upper=\strut},
  title=#2,#1
}

\newcommand{\eg}{\hbox{\emph{e.g.,}}\xspace}
\newcommand{\ie}{\hbox{\emph{i.e.,}}\xspace}
\newcommand{\etc}{\hbox{\emph{etc.}}\xspace}

\newcommand{\tool}{\textsc{AutoSC}\xspace}
\newcolumntype{L}[1]{>{\raggedright\arraybackslash}p{#1}}

\newtheoremstyle{sltheorem}
{}                
{}                
{}        
{}                
{\bfseries}       
{.}               
{ }               
{}                
\theoremstyle{sltheorem}
\newtheorem{definition}{Definition}

\newcommand{\code}[1]{{{\em \small\textsf{#1}}}}
\newcommand{\coderule}[1]{{\footnotesize\textsf{#1}}}
 \definecolor{dkgreen}{rgb}{0,0.6,0}
\definecolor{gray}{rgb}{0.5,0.5,0.5}
\definecolor{mauve}{rgb}{0.58,0,0.82}
\ifCLASSINFOpdf
\else
\fi

\begin{document}
%

\title{Combining Program Analysis and Statistical Language Model for Code Statement Completion}


\author{\IEEEauthorblockN{Son Nguyen and Tien N. Nguyen}
\IEEEauthorblockA{Computer Science Department\\
The University of Texas at Dallas, USA\\
Email: \{sonnguyen,tien.n.nguyen\}@utdallas.edu}
\and
\IEEEauthorblockN{Yi Li and Shaohua Wang}
\IEEEauthorblockA{Department of Informatics\\
New Jersey Institute of Technology, USA\\
Email: \{yl622,davidsw\}@njit.edu}
}


%


\maketitle



\input{abstract}


%
\IEEEpeerreviewmaketitle

\input{intro}

\input{problem_motivation}

\input{representation}

\input{approach}
\input{methodology}

\input{results}

\input{related}

\input{conclusion}


\section*{Acknowledgment}
This work was supported in part by the US National Science
Foundation (NSF) grants CCF-1723215, CCF-1723432, TWC-1723198,
CCF-1518897, and CNS-1513263.



%



\balance

\bibliographystyle{plain}
\bibliography{references,slamc,grapacc,dnn4c}

\end{document}

%% file: abstract.tex
\begin{abstract}
Automatic code completion helps improve developers' productivity in
their programming tasks. A program~contains instructions expressed
via code statements, which are~considered as the basic units of
program execution.
In this paper,~we introduce {\tool}, which combines program analysis
and the principle of software naturalness to fill in a partially
completed statement. {\tool} benefits from the strengths of both
directions, in which the completed code statement is both frequent~and
valid.
{\tool} is first trained on a large code corpus to derive the
templates of candidate statements. Then, it uses program analysis to
validate and concretize the templates into syntactically and
type-valid candidate statements. Finally, these candidates are ranked
by using a language model trained on the lexical form of the source
code in the code corpus.
Our empirical evaluation on the large datasets of real-world projects
shows that {\tool} achieves 38.9--41.3\% top-1 accuracy and
48.2--50.1\% top-5 accuracy in statement completion. It also
outperforms a state-of-the-art approach from 9X--69X in top-1
accuracy.
\end{abstract}

%% file: intro.tex
\section{Introduction}
\label{intro-section}

Code completion tool helps improve developers' productivity by
filling in the code during their editing. A program~contains
instructions in source code to~perform certain tasks. The procedure to
achieve a task is expressed via program statements, each of which is
considered as the basic unit of execution in a program. A statement
can declare a variable, define an expression, perform a simple action
by calling a method, control the execution flow of other
statements, create an object, or assign a value to a variable,
attribute, or field \cite{wiki-csharp}. Thus, in this work, we aim to
support automated code completion to help developers fill in their
current statements. During writing the body of a method, if a
developer finishes one or more tokens of the current statement, the
tool as requested will fill in the remaining tokens of that
statement. If (s)he finishes a statement, the tool will suggest
the entire next statement ({\em next-statement completion}). Let us
call it a
{\em statement completion} (SC) tool. SC encompasses next-statement
completion.


To build an effective and efficient SC tool, one would face the
following key challenges. First, the tool must predict the statement
that a developer intends to type next to perform the programming task
at hand. Second, the resulting code after completion must conform to
the syntactic and semantic constraints defined by the
programming language in use.

To address the first challenge, one can rely on the principle of
software naturalness~\cite{prem_naturalness}. Source code is naturally
written with certain regularity, \ie it is repetitive and does not
occur randomly. The code elements appear together because they are
intended by developers to achieve a programming task(s). Hindle {\em
  et al.}~\cite{prem_naturalness} showed that such regularity in
source code can be captured by statistical language models (LMs), \eg
$n$-gram model~\cite{ngram-wiki} can be leveraged to support code
completion for the next token. Thus, one could train an LM with a
large code corpus and use it to predict each token at a time until a
complete statement is suggested. However, the frequent code fragments
learned from different contexts might make the code after completion
syntactically or semantically incorrect. For example, after {\code{``i
  =''}, if the most frequent variable in a corpus is \code{i}, the
resulting code is {\code{``i = i;''}, which is invalid.
A naive solution that uses program analysis (PA) to enforce the
constraints in such output with multiple tokens would face
combinatorial explosion. For example, assume that at each step, a
model predicts and maintains $n$ most likely valid tokens, the number
of statements with $m$ code tokens is $n^{m}$.

To address the second challenge, an SC tool can apply PA with
program constraints on the candidate statements to eliminate the
invalid ones, the number of the remaining, valid candidates
is still large. The accuracy~of~such a naive solution is very low
due to the confounding effect of the accuracy of a prediction model
for each token (see Section~\ref{empirical}).
Another solution to this issue is to search for an entire code
statement. However, it is ineffective since statements are
project-specific and do not repeat often across different methods or
projects as reported in PCC~\cite{ase_17}. In fact, learning to
suggest {\em entire} statements is less effective than an SC tool that
is capable of filling the remaining token(s) of the current
statement.

This paper proposes {\tool}, which combines program analysis and
statistical LM in the process of statement completion. We aim to
benefit from the strengths of both directions in which LM produces
natural code sequences and PA enforces syntactic and type constraints.
{\tool} works in three phases.

First, it uses the $n$-gram LM on an abstraction level higher than
lexical code to learn to derive the most likely {\em candidate
  templates} for the current statement. A candidate statement is
modeled by a sequence of special annotations called {\em extended code
  tokens} ({\em excode} for short). An {\em excode} for a token is an
annotation representing the token type and/or data type, if
available. The token type encodes whether the token is a variable, a
field access, a method call, a type (class), \etc Such information in
a template helps a LM predict better the next token, \eg a variable
cannot be next to another. Data types help distinguish the code
fragments having the same meaning but with different variables' names,
\eg ``\code{int len = s.length();}'' and ``\code{int l =
  str.length();}'' have the same meaning of {\em ``Retrieving the
  length of a String and assign it to an int variable''}. With the
types instead of the variables' names, the two fragments have the same
template. Thus, {\tool} can learn templates from one place to suggest
for the other. Data~types also help distinguish the cases in which two
fragments with the same lexical tokens having different meaning. For
example, \code{x.next} means a variable \code{x} of a \code{Scanner}
accessing the field \code{next}, while in another place, it means a
variable \code{x} of a \code{LinkedList} calling the method
\code{next}. Data types thus help determine the accessible method
calls or field accesses for a variable.


At the second step, the candidate templates are syntactically and type
validated. Then, the valid templates are concretized into code
sequences. Finally,
all valid suggested code sequences are ranked based on their
occurrence likelihoods given the partial code. To do this, we train
another $n$-gram model on the lexical form of the source code in a code
corpus.


We conducted several experiments to evaluate {\tool} in statement
completion on a dataset used in the existing
approaches~\cite{prem_naturalness,fse13,saner18} with +460K statements
with a total of +1M suggestion points. Our results show that {\tool}
is very effective with top-1 accuracy of 40\% and top-5 accuracy of
49.4\% on average. That is, in 4 out of 10 cases, when a user requests
to complete his/her currently-written statement, (s)he can find the
remaining of the desired statement in the top of the
suggestion list.
Importantly, {\tool} significantly improves over the baseline model
using only $n$-gram on lexical code (up to {\bf 142X} in top-1
accuracy) and the model using lexical $n$-gram+PA (up to {\bf 117X} in
top-1 accuracy). It also improves over the state-of-the-art tool
PCC~\cite{ase_17} with {\bf 69X} higher in top-1 accuracy. In brief,
our contributions include


1. A model with PA+LM to complete the current
statement,

2. An empirical evaluation showing our model's
effectiveness and much better accuracy than the
state-of-the-art tool.












%% file: problem_motivation.tex
\section{Motivating Example}


\input{problem}

\input{motivation}

\input{overview}

%% file: problem.tex

Figure \ref{example1} partially shows a method in Apache
Ant~\cite{ant}.~Assume that the cursor is at line 5, right after
the \code{``=''} sign.~If a user requests a statement~completion (SC)
tool,~it will~complete the current statement, \ie the~assignment to
the variable \code{len}. A SC~tool would predict the intention of the
user and complete that assignment with the method
call \code{children.getLength()}. The tool suggests a ranked list of
candidate statements such as in Figure~\ref{example1}.
If the cursor is at the beginning of a statement, \eg at the beginning
of the line~5, the SC tool would suggest the entire assignment
statement, \eg \code{int len = children.getLength()} (next-statement
suggestion). That is, SC includes the functionality of next-statement
suggestion (\eg PCC~\cite{ase_17}).
Note~that, the SC tool is automatically invoked as the user
finishes typing a token in the middle of a statement, \eg after
\code{int}, \code{len}, \code{``=''},~\etc
%


To support statement completion, a model needs to consider the nature
of source code. Source code is strictly defined~by the syntax and
semantics of the programming language. Source code is also
repetitive \cite{prem_naturalness}. Thus, the methods for SC can~be
realized in the following: {\em information retrieval (IR) and pattern
mining, program analysis, and statistical language model}.

For {\em IR and pattern mining}, a model suggests to complete the
current statement by searching for the same/similar statement(s) that
have been seen in a corpus. When the retrieved statements have
occurred frequently, they can be viewed as code patterns. Such a
pattern or a retrieved statement can be used as the candidate for
completion. However, the tokens need to be filled for the current
statement might not be a pattern, leading to ineffectiveness of such
approach.
Moreover, while as single tokens, code is repetitive; as entire
statements, they are quite unique for specific projects. This
phenomenon was reported by Yang {\em et al.}~\cite{ase_17}. Indeed, in
our experiment (Section~\ref{results}), the portion of repeated
statements in our dataset is 25.9\%. That is, 3 out of 4 cases
on average cannot be correctly suggested by searching for the same
statements in the corpus of the previously-seen statements. As an
example, the statement \code{int len = children.getLength();} is not
used in any other project in our dataset. As an implication, to
suggest or complete a statement, a model cannot rely solely on
searching for the repeated statements as a whole.


%

%% file: motivation.tex
\begin{figure}[t]
\begin{center}
\begin{lstlisting}[captionpos=b, basicstyle={\footnotesize\ttfamily},numbersep=-4pt]
  NodeList listChildNodes(Node parent,NodeFilter f){
    NodeListImpl matches = new NodeListImpl();
    NodeList children = parent.getChildNodes();
    if (children != null) {
        int len = 	children.getLength();
	          	  	children.getLength();
	          		parent.numChildren();
	          		0;
\end{lstlisting}
\vspace{-0.15in}
\caption{A partial method in class \texttt{DOMUtil} of Apache Ant}
\label{example1}
\end{center}
\end{figure}

For the \textit{program analysis (PA)} direction, although the number
of valid candidates for the next token is limited, the number of
possible valid (complete) statements at the suggestion point might
be combinatorially explosive or even infinite. For the right side of
the assignment at line 5, the valid next-token candidates include the
appropriate prefix operators (\eg \code{``++''} and \code{``$--$''}),
the open parenthesis, field access, method call, and local variable
(\eg \code{children}, \code{f}, \etc). However, there is an infinite
number of valid statements at that point.
%
In brief, program analysis direction could produce a large number of
candidates with equal occurrence likelihoods, despite that the
candidates are syntactically or semantically~valid.



The \textit{statistical language models (LM)} leverage the fact that
code is highly repetitive and predictable~\cite{prem_naturalness}. The
next tokens to be filled are based on the frequent sequences of tokens
and the partial code.
Solely relying on those to fill in a statement, a model could face the
following issues. The first issue is caused by the fact that {\em the
  code in different places with the same lexical code sequence have
  different meaning}. For example, in one place, \code{x.next} means
the variable \code{x} of a \code{Scanner} in JDK accessing the field
\code{next}, while in another place, it means the variable \code{x} of
a \code{LinkedList} calling the method \code{next}. In this case, a LM
can mistakenly use one to suggest for another, \eg it could recommend
\code{``()''} after \code{x.next} for the field access of a
\code{Scanner}, which results in a semantic error.
Second, in contrast, to {\em represent the same meaning in different
  places in the same or different projects, one could use different
  names of the variables}. For example, the statement at line 5 is a
code fragment that performs the task of {\em ``retrieving the size
  (length) of a list of nodes''}. In other places,
we might see \code{int size = children.getLength()}. Those two code
fragments might be deemed as not performing the same task if
only the lexical tokens
are considered. Thus, an LM cannot learn from one
place to complete the statement in the other place.


Third, the names of method calls and field accesses might not appear
in the training data, leading to the {\em out-of-vocabulary} (OOV)
issue. This also applies to local variables due to their
method-specific nature. In natural language, human can understand a
sentence even an OOV word is missing. However, OOV could cause the
code un-compilable. Finally, even OOV does not occur, the completed
code by a LM could violate syntactic and semantic constraints. At line
5, the most likely next sequences of tokens include \code{i;},
\code{(}, or \code{)}, which are frequent in a corpus.  That would
induce {\em an ``undeclared variable''~error}.


%
%



\textit{From the above discussion, it is natural to combine PA~and LM
  to benefit from the strengths of both directions in completing the
  current statement}. For example, PA can be used to derive/select
the syntactically and type-valid candidate~statements from
the list produced by statistical LM, while the latter can apply
the principle of code repetition~\cite{prem_naturalness} to rank the
valid and most likely statements higher in the candidate list.

A naive LM+PA solution would use a statistical LM~to~predict the next
token one by one, and then use PA to filter~out the invalid ones and
rank the remaining ones according to their occurrence
likelihoods. However, doing so, the number~of valid
statements is still large.
%
%
Our experiment (Section~\ref{results}) showed that among those valid
ones, the correct one is rarely in the top 5 most likely candidates:
top-5 accuracy is $\approx$ 0.83\%.

%% file: overview.tex
\section{Key Ideas and Approach Overview}

We develop {\tool}, which combines program syntax and type
constraints and the naturalness principle of source
code~\cite{prem_naturalness} in the process of code statement
completion.
First, we use an LM on an abstraction level higher than lexical source
code to learn to derive the most likely candidate~templates for the
current statement.
%
%
Second, the candidate~templates are syntactically and type validated,
and concretized into one or more code sequence candidates. After all, 
we rank the candidate code statements accordingly to their occurrence 
likelihoods by another LM trained on lexical source code.
%

To overcome the issues of a LM on OOV and capturing high-level
abstraction of source code, we design a {\em template} as {\em a
  sequence of extended annotation code tokens} ({\em excode} for
short). An \textit{excode} for a token is an annotation representing
the token type and/or data type, if available (details in
Section~\ref{representation-section}).
For an identifier, its {\em excode} captures its token type, \ie a
variable, a field access, a method call, a type (class), \etc Token
types in a template helps a LM predict better the next token, \eg an
\code{'('} is needed after a method call. {\em excode}
also captures the data type if available. For example, \code{children}
is of the type \code{NodeList} at line 5.  The data type facilitates a
model to restrict possible method calls or field accesses. However,
the variable names are not kept in an {\em excode} because we want to
capture the code pattern at a higher level. In contrast, {\em excode}
keeps the name of the class that is declared (\eg \code{NodeList} in
\code{NodeList children}), the method that is called (\eg
\code{getLength}), the field that is accessed (\eg \code{next}). The
rationale is that those elements are designed to be (re)used in
different classes, methods in the same or different projects (\eg
libraries/frameworks). Such reused names would be useful for a model
to learn to apply in different places. The literals are not kept
because they tend to be project-specific except if they are special
literals, such as \code{null} or \code{0}. The other kinds of tokens
are kept intact.

These treatments help {\tool} learn better the candidate templates. At
line 5, the template has the left-hand~side of \code{TYPE(int)}
  \code{VAR(int)}, \code{OP(ASSIGN)}, and the right-hand side
  of \code{VAR(NodeList)} \code{OP(ACC)} \code{CALL} \code{(NodeList,}
  \code{getLength,} \code{0,int)} \code{LP RP}.
%
%
By raising the abstraction from the code, we aim to increase the
regularity/repetition to help a LM learn from other places to better
find the statement templates. For example, while the fragment
\code{len = children.getLength()} has never appeared in the project,
the above template occurs 6 times.

Our process of learning templates and concretizing into code helps our
model overcome OOV and the nature of locally-used variable names. The
templates at higher level are learned from one place and applied to
another, and PA is used to concretize them with concrete accessible
variables at the new place. The step of learning at template level
helps {\tool} cover more candidates (improving recall), while the use
of PA helps retain more valid ones (improving precision).

To enforce syntax and type constraints, we train an LM with the
sequences of {\em excode} to learn the statement templates, and use
that LM to suggest each {\em excode} by {\em excode} to form the
candidate templates. During that, syntactical and type rules
are applied to those candidate templates to enforce their validity.

The second LM on lexical source code at the last step helps select the
variable names when there still exist multiple candidates of code
sequences. When several valid variables are valid, the lexical LM
selects the names that come naturally and frequently at the place. For
example, the tokens \code{children}, \code{node}, \code{parent}, \etc
often go together. Thus, at line 5, the variable name \code{children}
likely occur than \code{student}, \code{network}, \etc

%% file: representation.tex
\section{Extended Code Annotation}
\label{representation-section}
  


\input{design}

\input{code_annotation}

%% file: design.tex
\subsection{Design Strategies}

\input{annotation_rules}

We present \textit{extended code annotation (excode)}, a code
representation designed for \textit{SC}. Let us explain what
information needs to be encoded.
%
%
We first aim to encode \textbf{token type} of~a code token. That is,
we need to encode whether a code token is a keyword, separator,
operator, method call, field access, variable, \etc This
enables {\tool} to learn program syntaxes on the validity of a next
code token, \eg \textit{``A left parenthesis must appear after the
  method call \code{next}, not after the field \code{next}''}.
%
%
Additionally, for the validation of the type constraints, the
\textbf{data type} of code tokens, especially of method call, field
access, and variable, also needs to be encoded. For example, the RHS
expression of the assignment at line 5 must be of the type \code{int}
or \code{Integer} because the LHS variable is of the type
\code{int}.
%
%



Because local variables are used locally, their names and
meaning might be different in different methods. Thus, they can not
be learned by a LM in a method to apply to the local variables
but with different variable names in other methods.
%
Thus, the names of local variables should be abstracted in the
representation to better capture code regularity. Meanwhile, the
names of data types, methods, and fields are kept
since those elements are designed to be reused in other
places. Thus, those names can be learned from one place and be applied
to~others.

%% file: annotation_rules.tex
\begin{table*}[t]
\centering
\footnotesize
\tabcolsep 2pt
\caption{{\em Excode} Annotation Rules for Code Tokens}
\label{annotation_rules}
\begin{tabular}{lll}
\toprule
Token Role      		& Construction Rule                                    			& Example code$\to$\textit{excode} 	\\ \hline
Keyword         		& To corresponding reserved token                     			& \code{if}$\to$\code{IF}, \code{for}$\to$\code{FOR}         				\\ 
Operator $o$      	& OP{(}$name(o)${)}                                      			& \code{.}$\to$\code{OP(ACC)}, \code{=}$\to$\code{OP(ASSIGN)}        				\\ 
Separator       		& To corresponding reserved token                      			& \code{(}$\to$\code{LP}, \code{)}$\to$\code{RP}        				\\ 
Data type $T$     	& TYPE{(}$T${)}                                          			& \code{int}$\to$\code{TYPE(int)}, \code{String}$\to$\code{TYPE(String)}   \\ 
Variable $v$      	& VAR{(}$type(v)${)}                                  			& \code{len (int)}$\to$\code{VAR(int)}, \code{parent (Unknown)}$\to$\code{VAR(Unk)}         				\\ 
Literal $l$       	& LIT{(}$littype(l)${)}                                  			& \code{"hello"}$\to$\code{LIT(String)}, \code{123}$\to$\code{LIT(int)}     \\ 
Method call $m$  	& CALL{(}$Type(m)$,$name(m)$,$argcount(m)$,$rt(m)${)} 	& \code{subString(1)}$\to$\code{CALL(String,subString,1,String) LP LIT(int) RP}   		   			\\ 
Field access $f$  	& FIELD{(}$Type(f)$, $name(f)$, $type(f)${)}         				& \code{node.name}$\to$\code{VAR(Node) OP(ACC) FIELD(Node,parent,String)}        				\\ 
Special literal 		& To corresponding reserved token                      			& \code{null}$\to$\code{NULL}, \code{0}$\to$\code{ZERO}, \code{""}$\to$\code{EMPTY}          				\\ \toprule
\end{tabular}
\end{table*}

%% file: code_annotation.tex
\subsection{Extended Code Tokens Annotation and Concretization}

\begin{definition}[{\em Token Type}]
The token types in a program with regard to a programming language
include keyword,~operator, separator, data type, method
call, field, variable, and~literal.
\end{definition}

For \code{children.getLength()}, the token types of
\code{children} and \code{getLength} are variable and method call, respectively,
while \code{\bf.} (access) is an operator, and \code{(} and
\code{)} are separators \code{LP}~and~\code{RP}.

\begin{definition}[{\em Excode Token}]
An {\em excode} token is an annotation corresponding to a code token,
that represents its syntactic and type information, including its
{\em token type} and {\em data type}.
\end{definition}

Table \ref{annotation_rules} shows the rules to construct {\em excode}
tokens for popular kinds of code tokens. For \code{children} in
\code{children.getLength()}, which has the role of a variable, its
corresponding {\em excode} token consists of the annotations
``\code{VAR}'', ``\code{(}'', its data type \code{NodeList}, and
"\code{)}''. For method calls and field accesses, the information
including the enclosing type name, return type, and the arguments, are
additionally incorporated in the {\em excode} tokens. For example, the
{\em excode} of \code{getLength} in \code{children.getLength()} is
\code{CALL(NodeList,getLength,0,int)}.

\begin{definition}[{\em Excode} {\em annotation function $\alpha$}]
  \label{alpha}
The annotation function $\alpha(C)$ on a code sequence
$C=c_1c_2...c_n$, defines the corresponding \textit{excode} sequence
$E=e_1e_2...e_n$, such that $e_i$ is the corresponding
\textit{excode} token of $c_i$ defined in Table~\ref{annotation_rules}.
\end{definition}

Since $C$ is the current partial code, to realize $\alpha$, we
perform partial program analysis using {\em PPA}~\cite{ppa08} to get
token types and data types in a best-effort fashion.

\begin{definition}[{\em Excode} {\em token concretization function}]
\label{def_pi}
The concretization function $\pi(e,V)$ on an \textit{excode}
token $e$ and the set $V$ of the accessible variables and
 fields of the current class of the method, defines the set of
code tokens as follows:
\[\pi(e,V)=
  \begin{cases}
   \{v:v \in V, type(v) = type(e) \}   & \text{if } e \text{ is a variable}\\
   \{c\} & \text{otherwise}
  \end{cases}
\]
where, $c$ is the respective non-variable token listed in
Table~\ref{annotation_rules}.
\end{definition}

In Figure~\ref{example1}, $\pi('$\code{VAR(NodeList)}$',V)$ =
$\{$\code{children}$\}$, where $V$ contains the set of accessible
(global/local) variables of the method \code{listChildNodes}. Note
that, literals will not be~concretized except if they are special
literals, such as \code{null} or~\code{0}.

\begin{definition}[{\em Excode} {\em sequence concretization function}]
The sequence concretization function $\Pi(E,V)$ on an {\em excode}
sequence of length $n$, $E_n=e_1e_2...e_n$, in a method and the set of
the method's accessible variables and fields $V$, defines a set of code
sequences of length $n$, in which each code sequence
$C_n=c_1c_2...c_n$, $c_i \in \pi(e_i, V)$, for $\forall i \in [1,n]$.
\end{definition}

\begin{definition}[\textit{Excode} {\em expression}]
In a method having the set of accessible variables $V$, an {\em
  excode} expression $expr$ is an {\em excode} sequence with one or
more {\em excode} tokens, such that there is at least one code
sequence $C$ in $\Pi(expr, V)$ that is a valid code expression
according to the programming language.
\end{definition}

In our example, the \textit{excode} sequence \code{VAR(NodeList)}
\code{OP(notEquals)} \code{NULL} is an {\em excode} expression since
there exists a concretization to obtain a valid expression \code{children !=
  null}.


\begin{definition}[{\em Excode} {\em statement}]
In a method having the set of accessible variables $V$, an {\em excode}
statement $stm$ is an {\em excode} sequence with one or more {\em
  excodes}, such that there is at~least one code sequence $C$ in
$\Pi(stm, V)$ that is a valid statement.
\end{definition}


We use {\em excodes} to represent a statement template.
%
For example, \code{TYPE(int)} \code{VAR(int)}
\code{OP(ASSIGN)} \code{VAR(NodeList)} \code{OP(ACC)}
\code{CALL(}\code{NodeList},\code{getLength},\code{0,int)} \code{LP} \code{RP}
is a template.

%% file: approach.tex


\input{complete_template}

\input{realize_template}

\input{rank_code}


%% file: complete_template.tex
\section{Identifying Candidate Templates}

Given the partial code $P$, {\tool} first parses $P$ to~build the {\em
  excode} sequence $E=e_1e_2...e_n$.
It uses the $n$-gram model~\cite{ngram-wiki} that is trained on the
{\em excode} sequences built from a code corpus to predict each {\em
  excode} one by one that most likely follows $E$. It also uses
rules for program constraints to derive the valid candidates of {\em
  excode} tokens and sequences.~The~resulting {\em excode} sequences
represent templates.
Let us detail it.




\subsection{Training $n$-gram LM with excodes to predict next~excode}



To predict the next {\em excode}, any statistical LM is
applicable~\cite{white-msr15,saner18,dam16,tu-fse14}. Without loss of
generality, we use $n$-gram LM~\cite{ngram-wiki}. The model is
trained on the {\em excode}~sequences built from a corpus.
%
For prediction, given $E$
and an {\em excode} candidate $\epsilon$, the likelihood, that
$\epsilon$ is the next {\em excode} token following $E$, is estimated
using the trained $n$-gram LM,~$\phi_{excode}$:
\indent \indent \indent \indent \indent $P(\epsilon|E) = \phi_{excode}(e_1e_2...e_n\epsilon) \quad \quad \quad (1)$

\subsection{Deriving the next excode sequence for statement template}

Next, using $\phi_{excode}$, {\tool} identifies the most likely valid
{\em excode} one at a time, and then composes them to obtain
the candidates for statement template.
%
%
Specifically, Algorithm~\ref{agl1} shows how {\tool} identifies
candidate templates. In this algorithm, the partial code is first
parsed into the corresponding \textit{excode} sequence (line 2). The
next sequences are suggested by expanding the {\em excode sequence}
token-by-token until encountering the end-statement token ``\code{;}''
or the length of the expanded sequence reaches the pre-defined maximum
length of code statements (lines 3, 7--9). For each expansion step,
{\tool} applies the syntax rules and accessibility rules (will be
explained later) to enforce program constraints. The set of {\em
  valid} candidates of the next {\em excode} token is stored in
$\mathbb{C}$. Then, it selects the top $K$ (predefined value)
most likely tokens (line 11).
These {\em excodes} are concatenated with $E$ to
form new candidates that are recursively expanded (lines
13--16).

\input{algorithm}


\subsection{Enforcing syntax rules and accessibility rules to
  decide the candidates for the next {\em excode} token}


A vocabulary $\mathscr{V}$ is a set of all distinct
\textit{excode} tokens. Since code has strict syntax and semantics,
for \textit{excode} sequence $E$, the valid next {\em excode} token
following $E$
is restricted by program constraints/rules: \textit{Syntax rules} and
\textit{Accessibility rules}.

\begin{definition} [{\em Program syntax rule}]
Given the {\em excode sequence} $E$ = $e_1e_2...e_n$, the vocabulary
$\mathscr{V}$ of all {\em excode} tokens, a program syntax rule
$r_{syntax}$ when applying on $E$ will return a set $\mathscr{S}$ of
{\em excode} tokens in the vocabulary such that the resulting {\em
  excode} sequence $E'=e_1e_2...e_n\epsilon$ does not violate a 
syntax rule of a programming language.
Mathematically, a program syntax rule $r_{syntax}$ is a relation $r:
(\mathscr{V})^{*} \to 2^{\mathscr{V}}$, $r_{syntax}(E) = \mathscr{S}$,
where $\mathscr{S} \subseteq \mathscr{V}$ is the set of tokens,
such~that $\forall \epsilon \in \mathscr{S}$, 
$E'=e_1e_2...e_n\epsilon $ does not violate a syntax rule.
 
\end{definition}

For example, the code \code{int len =} has the {\em excode} sequence
of \code{TYPE(int) VAR(int) OP(ASSIGN)}. The {\em excode} tokens
\code{OP(ASSIGN)} and \code{OP(ACC)} are excluded from
$r_{syntax}(E)$ because an \code{``=''} or \code{``.''} cannot occur
after the \code{``=''} sign.
In this example, $r_{syntax}(E)$ includes literal, variable, method
call field access, data type, prefix operators, or open parenthesis.
Note that to check for $\epsilon$, instead of checking all syntax
rules on $E'=e_1e_2...e_n\epsilon$ at each expansion step, for
efficiency, we could check the validity of $\epsilon$ based on the
last token $e_n$, and finally, check on the syntactic validity of
entire sequence at the last step when the end of statement is reached.

\begin{definition} [{\em Accessibility rule}]
Given the {\em excode sequence} $E = e_1e_2...e_n$, the vocabulary
$\mathscr{V}$ of all {\em excode} tokens, an accessibility rule
$r_{access}$ when applying on $E$ will return a set $\mathscr{A}
\subseteq \mathscr{V}$ of the {\em excode} tokens in the vocabulary that are
accessible at the current state of $E$. That is, $r_{access}$ is a
relation $r_{access}: (\mathscr{V})^{*} \to 2^{\mathscr{V}}$,
$r_{access}(E) = \mathscr{A}$ such that $\mathscr{A}$ includes the
\textit{excode} tokens which correspond to the following cases:


1) All declared local variables within the current scope are
  valid. In Figure~\ref{example1}, accessible local
  variables are \code{VAR(Node), VAR(NodeFilter), VAR(NodeListImpl)}
  and \code{VAR(NodeList)}.
 
2) All the accesses to the fields and the calls to the methods in
  the enclosing class are accessible.

3) The accessible field accesses and method calls of a
  variable. For example, for a sequence $E$ ending with
  \code{VAR(NodeList) OP(ACC)}, all accessible field accesses and
  method calls in \code{NodeList} are included in $r_{access}(E)$.

4) All data types and literals are valid.

5) All keywords, separators, and operators are valid.


 




\end{definition}

\begin{definition} [{\em Valid next \textit{excode} token}]
\label{validtoken}
For a sequence, a \textit{excode} token is considered as valid if it
satisfies all \textit{Syntax rules} and \textit{Accessibility
  rules}. That is, given an \textit{excode} sequence $E = e_1e_2...e_n$,
the set of valid candidates $\mathbb{C}$ is $r_{syntax}(E) \cap
r_{access}(E)$ for all syntax rules and accessibility rules.
\end{definition}



%% file: algorithm.tex
\begin{algorithm}[t]
\caption{Identifying Candidate Templates}
 \label{agl1}
\begin{algorithmic}[1]
\Function {identifyTemplates}{$partialCode$, $project$}
	\State $E = \alpha(partialCode)$  \Comment{Def.~\ref{alpha}}
	\State $genSeqs = expandExcodeSeq(E, project)$
	\State $templs=extractRemainingParts(genSeqs,E)$
	\State \textbf{return} $templs$
\EndFunction

\Statex

\Function {expandExcodeSeq}{$exSeq$, $proj$}
	\If {$isEnded(exSeq) \lor reachMaxLen(exSeq)$}
		\State \textbf{return} $\{exSeq\}$
	\EndIf
	\State $\mathbb{C} = getValidNextToken(exSeq, proj)$ \Comment{Def.~\ref{validtoken}}
	\If {$\mathbb{C} = \emptyset$} \textbf{return} $\emptyset$
	\EndIf
	\State $topCands = rank(\mathbb{C}, exSeq, \phi_{excode}, K)$ \Comment{Form. 1}
	\State $exSequences = \emptyset$
	\ForAll {$cand \in topCands$}
		\State $newSeq = concat(exSeq,cand)$
		\State $newTempls = expandExcodeSeq(newSeq,proj)$
		\State $exSequences.addsAll(newTempls)$
	\EndFor
	\State \textbf{return} $exSequences$
\EndFunction
\end{algorithmic}
\end{algorithm}

%% file: realize_template.tex
\section{Validating Candidate Templates}
\label{validate_template}

Fully semantic checking with respect to the current programming
language (\eg Java) is always desired. However, it is impossible to
do so for the candidate templates, which are expressed as the
sequences of {\em excode} tokens and do not contain concrete lexemes
of variables. Because our design is to have {\em excodes} contain data
type information, we focus on performing type checking. With
type checking, we can eliminate a large number of templates with
incorrect and inconsistent types.


In general, one could use a type checker for the current programming
language, \eg Java type checker. However, we are dealing with
partially complete code and there are potentially program entities
whose types cannot be resolved by PPA~\cite{ppa08}. In those cases,
the variables without type information are annotated with
\code{Unknown} type. Thus, we build a type~checker for {\em
  excode} with the accommodation of the \code{Unknown} type.


\input{type-rules}

{\tool} performs type inference at the same time as type checking on
{\em excode} statements and expressions using the rules in
Table~\ref{tab:rules}.
%
The process of type checking is similar to type checking for the
source code in Java. However, there are two key differences. First,
it works at the {\em excode} statements/expressions corresponding
to the statements/expressions at the source code level (note:
variables' names are not there). Second, due to unresolvable
types, {\tool} has to consider \code{Unknown} type in a flexible
manner, \eg that type does not violate any subtype constraint. Let us
explain the key type-checking rules.





{\bf 1. Literal.} When seeing the {\em excode} \code{LIT(T)} that
represents a literal with a type \code{T}, we consider \code{T}
as its type.

{\bf 2. Variable.} When seeing a \code{VAR}, if the type
of {\em excode} is available, we use it. Otherwise, the resulting
type is \code{Unknown}.

{\bf 3. Assignment.} The LHS and RHS expressions are type-checked first.
If both types are known, the type of RHS must be a subtype or equal to
the type of LHS. If either of them are \code{Unknown}, we consider
the assignment as valid with the known type.
If both are \code{Unknown}, the resulting type is~\code{Unknown}.

{\bf 4. Prefix.} If the operator is a negation
and if the type of $e$ is available, it must be \code{boolean},
otherwise, it must be convertible to a numeric type (\code{char, short, int}, \etc) The resulting type is \code{boolean} or
a numeric type, accordingly. If the type of $e$~is \code{Unknown}, the
result depends only on the operator (Table~\ref{tab:rules}).

{\bf 5. Postfix.} The type of $e$ must be convertible to a numeric
type or it is unavailable. The resulting type is numeric.


{\bf 6. Comparison.} The type of one side must be a sub-type or equal
to the type of the other side, or the type of at least one of them
must be \code{Unknown}. The resulting type is \code{boolean}.


{\bf 7. Infix.} Both expressions on two sides need to be
type-checked. If both types are not \code{Unknown}, the type of one
side must be a subtype or equal to the other, and the expression
is assigned with the super type. If the type of one of the
two sides is \code{Unknown}, the expression is assigned of the type of
the known one. Otherwise, the type of the expression is
\code{Unknown}.


{\bf 8. Method Call.} The expressions for the receiver and the
arguments need to be type-checked first. The type of each argument (if
available) must be a subtype or equal to the type of the corresponding
formal parameter in the declaration of the method. The return type is
used as the type of the call.

{\bf 9. Constructor Call.} A constructor call is handled similarly as
a method call except that the declared type is used and the method
name is the same as the class name.

{\bf 10. Field Access.} The receiver needs to be type checked. The
class of the field must be the same as the respective type stored in
the {\em excode}.

{\bf 11. Variable Declaration.} The RHS expression (if any) needs to
be type checked, and its type (if available) must be a subtype or
equal to the type stored in the {\em excode} \code{VAR(T)}.


{\bf 12. For/While/If statement.} The components in the {\em excode}
of such a statement need to be type checked. The
conditional control statement must be of the type \code{boolean}
or \code{Unknown}.

{\bf 13. Expression/Block Statement.} Each statement in each of
those compound statements needs to be type checked.

{\bf 14. Return statement.} The expression needs to be type-checked
and its type must be a subtype or equal to the return type of the
enclosing method.

\begin{definition} [{\em Type-correct candidate template}]

  Given an {\em excode} sequence $E$ representing the current partial
  code, the template $T$ (as an {\em excode} sequence) is considered as a
  type-correct candidate template if the sequence concatenated by
  $E$ and $T$ is type checked by our rules.
%
\end{definition}

In Figure \ref{example1}, both candidates \code{0} and 
\code{VAR(NodeList)} \code{OP(ACC)} \code{CALL}\code{(NodeList,getLength,0,int) LP}
\code{RP} are type-correct.



%% file: type-rules.tex
\begin{table}[t]
  \centering
  \footnotesize
  \tabcolsep 2pt
  \caption{Key Type Check Rules for {\em Excode} Sequences}
  \vspace{-0.10in}
    \begin{tabular}{l|c|l}
    \hline
    \cellcolor{lightgray} {\bf Syntax} & \cellcolor{lightgray} {\bf Type} &\cellcolor{lightgray} {\bf Type Check ($e$: \coderule{T})} \\
    \hline
    \cellcolor{lightgray} {\bf Excode Seqs} & \cellcolor{lightgray} {\bf of $E$} & \\
    \hline
    Literal    & & \cellcolor{lightgray}  \\
    $E$ ::= \coderule{LIT(T)}  & \coderule{T} & \eg  \coderule{LIT(String)}: \coderule{String} \\
    \hline
    Variable   & & \cellcolor{lightgray}   \\
    $E$ ::= \coderule{VAR(T)}  & \coderule{T} & \eg  \coderule{VAR(int): int} or \coderule{VAR(Unk): Unk}\\
     \hline
    Assignment   & & \cellcolor{lightgray} $e_1$:\coderule{T1}, $e_2$:\coderule{T2}  \\
    $E$ ::= $e_1$ \coderule{OP(=)} $e_2$ & \coderule{T1} & \coderule{T2} $\subseteq$ \coderule{T1} if (\coderule{T1} $!=$ \coderule{Unk}) and (\coderule{T2} $!=$ \coderule{Unk}) \\
    & \coderule{T1} & else if \coderule{T1}!=\coderule{Unk}\\
    & \coderule{T2} & else if \coderule{T2}!=\coderule{Unk}\\
    & \coderule{Unk} & if \coderule{T1} = \coderule{T2} = \coderule{Unk}\\
    \hline
    Prefix op   & & \cellcolor{lightgray} $e$: \coderule{T} \\
    $E$ ::= \coderule{OP(op)} $e$ & \coderule{bool} & if op = \coderule{not} and ((\coderule{T}=\coderule{bool}) or (\coderule{T}=\coderule{Unk})) \\
    & \coderule{num} & otherwise ((\coderule{T}= \coderule{num}) or (\coderule{T}=\coderule{Unk}))   \\
& & \coderule{num} {\em types include} \coderule{char},\coderule{short}, \coderule{int}, etc.\\
    \hline
    Postfix op   & & \cellcolor{lightgray} $e$: \coderule{T} \\
    $E$ ::= $e$ \coderule{OP(op)} & \coderule{num} & if ((\coderule{T}= \coderule{num}) or (\coderule{T}=\coderule{Unk})) \\
    & & \coderule{num} {\em types include} \coderule{char},\coderule{short}, \coderule{int}, etc.\\
    \hline
    Comparison   &  & \cellcolor{lightgray} $e_1$: \coderule{T1}, $e_2$: \coderule{T2} \\
    $E$ ::=  & \coderule{bool} &  if (\coderule{T1} $\subseteq$ \coderule{T2}) or (\coderule{T2} $\subseteq$ \coderule{T1})\\
$e_1$ \coderule{OP(op)} $e_2$    &  &  or (\coderule{T1}=\coderule{Unk}) or (\coderule{T2}=\coderule{Unk}) \\
    \hline
    Infix   & & \cellcolor{lightgray} $e_1$: \coderule{T1}, $e_2$: \coderule{T2} \\
    $E$ ::=  & \coderule{T2} &  if (\coderule{T1} $\subseteq$ \coderule{T2}) != \coderule{Unk} \\
$e_1$ \coderule{OP(op)} $e_2$    & \coderule{T1} &  else if (\coderule{T2} $\subseteq$ \coderule{T1}) != \coderule{Unk}\\
     & \coderule{T1} & else if \coderule{T1} != \coderule{Unk} \\
    & \coderule{T2} & else if \coderule{T2} != \coderule{Unk} \\
    & \coderule{Unk} & else if both are \coderule{Unk}\\
    \hline
     Method Call  &  & \cellcolor{lightgray} $e$: \coderule{T}, $e_1$: \coderule{T1}, ..., $e_n$: \coderule{Tn}  \\
$E$ ::= $e$ \coderule{OP(ACC)}  & & decl: \coderule{RT T::m (T1' p1, T2' p2, ..., Tn' pn)} \\
\coderule{CALL(T,m,$n$,RT)} &  \coderule{RT} & if (\coderule{Ti} $\subseteq$ \coderule{Ti'}) or (\coderule{Ti}=\coderule{Unk}) for $i=1..n$ \\
LP $e_1$,..., $e_n$ RP & & \\
    \hline
    Constructor Call & & \cellcolor{lightgray} $e$: \coderule{T}, $e_1$: \coderule{T1}, ..., $e_n$: \coderule{Tn} \\
    $E$ ::=  &  & decl: \coderule{T T::T(T1' p1, T2' p2, ..., Tn' pn)}\\
 \coderule{CCALL(T,T,$n$,T)}  & \coderule{T} & if (\coderule{Ti} $\subseteq$ \coderule{Ti'}) or (\coderule{Ti}=\coderule{Unk}) for $i=1..n$\\
      \coderule{LP} $e_1$,...,$e_n$ \coderule{RP}    & & \\
    \hline
    Field Access    & & \cellcolor{lightgray}  $e$: \coderule{T} \\
    $E$::= $e$ \coderule{OP(ACC)}  & \coderule{FT} & \\
    \coderule{FIELD (T,f,FT)} & & \\
    \hline
    \hline
    \cellcolor{lightgray} {\bf Statement} & & \\
    \hline
    Variable Decl & & \cellcolor{lightgray} $e$: \coderule{T'}\\
    E ::= \coderule{VAR(T)}[=$e$], & \coderule{T} & if (\coderule{T'} $\subseteq$ \coderule{T}) or (\coderule{T}=\coderule{Unk})\\
    \hline
    ForStmt S::= & \coderule{void} & $i_1$: \coderule{Ti1}, ..., $i_n$: \coderule{Tin} \\
    for ($i_1$,...,$i_n$ ; $e$;  & & $e$: \coderule{T}, \coderule{T}=\coderule{bool} or \coderule{T}=\coderule{Unk},  \\
    $u_1$, ..., $u_m$) S1 & & $u_1$: \coderule{Tu1}, ..., $u_m$: \coderule{Tum}, $S1$: \coderule{T1} \\
    \hline
    S::= while (e) S1 & \coderule{void} & \cellcolor{lightgray}  $e$: \coderule{T}, \coderule{T}=\coderule{bool} or \coderule{T}=\coderule{Unk}, $S1$: \coderule{T1} \\
    \hline
    S::= if (e) S1 & & \cellcolor{lightgray}   $e$: \coderule{T}, \coderule{T}=\coderule{bool} or \coderule{T}=\coderule{Unk}, $S1$: \coderule{T1}, $S2$: \coderule{T2} \\
       $[$else S2$]$ & & \\
    \hline
    ExprStmt S::= e ; & \coderule{T} & \cellcolor{lightgray}  $e$: \coderule{T} \\
    \hline
    Block S::=$s_1$,.,$s_n$ & \coderule{void} & \cellcolor{lightgray}  $e_1$: \coderule{T1}, ..., $e_n$: \coderule{Tn}\\
    \hline
    Return S::= return $e$ & \coderule{T} & \cellcolor{lightgray}  $e$: \coderule{T}\\
    & & decl: \coderule{RT T::m (T1' p1, T2' p2, ..., Tn' pn)} \\
    & & \coderule{T} $\subseteq$ \coderule{RT} \\
    \bottomrule
    \end{tabular}%
  \label{tab:rules}%
\end{table}%

%% file: rank_code.tex
\section{Concretizing Statement Templates and Ranking Code Candidates}

This section describes how the type-correct candidate templates as
\textit{excode} sequences are converted to code candidate sequences
with the accessible variables in the current scope. The most
likely code sequences are ranked based on their occurrence
likelihoods computed by an LM. Let us detail it.

\input{realize_algo}



\noindent {\bf Concretization.} Algorithm~\ref{realize_algo}
shows our procedure. Each \textit{excode} token is converted into code
tokens using function $\pi$ (Def.~\ref{def_pi}).
These tokens are used to initiate a set of code~sequences (lines
9--12) or concatenated with the current concretized code sequences to
create the new ones (lines 14--17). The process recursively
continues until the end of the template.

\noindent {\bf Training an LM on lexical code and Ranking candidate
  statements.}
To rank the candidate code statements, we train an $n$-gram model,
$\phi_{lexemes}$, on the lexical forms of the source code in a 
corpus.
For training, all source files are tokenized based on naming
conventions (Camelcase and Hungarian), and the
obtained tokens are normalized to lowercase. The trained LM is used to
estimate the occurrence likelihoods of the code sequence that is
concatenated from the current code and the candidate statement.  That
is, given the current code $C$, the likelihood of the
candidate statement $\gamma$ is: $\phi_{lexemes}(concat(C_{lexemes},
\gamma_{lexemes}))$, where $C_{lexemes}$ and $\gamma_{lexemes}$ are
the lexical forms of $C$ and $\gamma$, respectively.


%% file: realize_algo.tex
\begin{algorithm}
\caption{Concretizing Candidate Template}
 \label{realize_algo}
\begin{algorithmic}[1]
\Function {concretize}{$templ$, $V$}
	\State $codeCands = concretizeNext(templ, V, \emptyset, 1)$
	\State \textbf{return} $codeCands$
\EndFunction
\Statex
\Function {ConcretizeNext}{$templ$,$V$,$currCands$,$i$}
	\If {$i > len(templ)$}
		\State \textbf{return} $currCands$
	\EndIf
	\State $codeCands =\emptyset$
	\State $codeTokens = \pi(templ[i], V)$ \Comment{Def.~\ref{def_pi}}
	\If {$currCands = \emptyset$}
		\ForAll {$t \in codeTokens$}
			\State $newCand = concat(EMPTY\_SEQ, t)$
			\State $codeCands.adds(newCand)$
		\EndFor
	\Else
		\ForAll {$t \in codeTokens$}
			\ForAll {$cand \in currCands$}
				\State $newCand = concat(cand, t)$
				\State $codeCands.adds(newCand)$
			\EndFor
		\EndFor
	\EndIf
	\State \textbf{return} $ConcretizeNext(templ,V,codeCands,i+1)$
\EndFunction
\end{algorithmic}
\end{algorithm}

%% file: methodology.tex
\section{Empirical Methodology}
\label{empirical}

We have conducted several experiments to empirically evaluate {\tool}
in statement completion. For that, we seek to answer the following
research questions:



\noindent\textbf{RQ1}: {\bf \textit{Accuracy and Comparison}}. How
accurate is {\tool} in {\em current statement completion} and {\em
  next statement suggestion}? How is it compared with the
state-of-the-art tool PCC~\cite{ase_17}?

\noindent\textbf{RQ2}: {\bf \textit{Intrinsic Accuracy}}. How accurate
is {\tool} in completing code statement on various factors including
code sequences' lengths and code token types?

\noindent\textbf{RQ3}: {\bf \textit{Sensitivity Analysis}}. How do
various factors affect our model, \eg completion position, thresholds,
and data's sizes?

\noindent\textbf{RQ4}: {\bf \textit{Time Complexity}}. What is our training/testing time?

\input{datasets}

\input{eval_setup}

%% file: datasets.tex
\subsection{Subject Systems}
\label{datasets}

In this study, we collected the same data set of Java projects used in
the existing studies in code completion~\cite{prem_naturalness,fse13}
(Table \ref{subject}) (\textit{Large Corpus}). In the dataset, the
average number of code tokens in a statement is 8.8, whereas more than
85\% of the code statements contain less than 12 code tokens.

For comparison on next-statement (NS) suggestion, we also used the
same dataset as in PCC~\cite{ase_17} (\textit{Small Corpus} in
Table~\ref{dataset2}). In the training data, 90\% of statements contain
less than 12 tokens. The test dataset contains only 10 individual
files without their projects. Both training and test data are much
smaller than our Large Corpus.

In our experiments, to balance between the completion effectiveness
and efficiency, we set the maximum number of tokens in a statement of
12.


\begin{table}[]
\centering
\caption{Large Corpus}
\label{subject}
\begin{tabular}{lllll}
\toprule
Project   &\#files	& \#statements & \#unique tokens & \begin{tabular}[c]{@{}l@{}}AVG \#tokens\\ in a statement\end{tabular} 	\\ \hline
Ant       &1,196			&50,041             & 3,721                &	8.5					\\
Batik     &1,657			&82,193             & 5,032                &	8.8					\\
Cassandra &687			&47,874             & 3,757                &	9.8					\\
Log4J     &309			&8,920              & 899                	  &	9.7					\\
Lucene    &3,681			&117,447            & 7,191                &	8.1					\\
Maven2    &378			&10,029             & 1,906                &	9.4					\\
Maven3    &850			&18,060             & 2,749                &	9.6					\\
Xalan-J   &958			&59,430             & 3,427                &	8.9					\\
Xerces    &829			&66,275             & 3,769                &	8.8					\\ \toprule
\end{tabular}
\end{table}

\begin{table}[]
\centering
\caption{Small Corpus}
\label{dataset2}
\begin{tabular}{l|l|l}
\toprule
                                                              	 & Training data 		& Test data \\ \hline
\#Files                                                   	 & 2,415      			& 10     			\\ 
\#Methods                                           	 & 6,680        		& 59      			\\ 
\#Statements                                         & 37,050        		& 440      			\\ 
\#Unique tokens                                    & 7,781        			& 304       			\\ 
AVG \#tokens in a statement 			 & 9.2        			& 6.2       			\\ 
\toprule
\end{tabular}
\end{table}

%% file: eval_setup.tex
\vspace{-0.06in}
\subsection{Evaluation Setup, Procedure, and Metrics}

We used the same setting with data across projects as in existing
work~\cite{prem_naturalness,tu-fse14,fse13}.  That is, we divided the
source files of a project into 10 equal folds. We performed 10-fold
cross-validation: each fold was chosen for testing, while the
remaining folds and other projects were used for training.

%

%

Accuracy on statement completion is measured as follows. For a method
in a source file in the test data, our evaluation tool traverses its
code sequentially from the beginning. At a position $i$ in a method
with a code sequence $M_n = c_1c_2...c_n$, a tool computes the top $k$
most likely code sequences, $s_1$, $s_2$,..., $s_k$, for the remaining
of the current statement based on the previous code sequence from the
start of the method to the position ($i-1$): $c_1c_2...c_{i-1}$.
If the actual code sequence, from $i$ to the end of the current
statement $s_i$ at the position $t$ is among the above $k$ suggested
sequences, we count this as a \textit{hit}. The top-$k$
accuracy is the ratio of the total hits over the number of tokens.
Top-$k$ accuracy for a project is computed on all positions of its
methods in cross validation.

%
%



Note that, for the compound statements
including \textit{if-then}, \textit{if-then-else}, \textit{switch}, \textit{for}, \textit{while},
and \textit{do-while} statements, we run a model to complete/suggest
their control expressions.
%
Moreover, at a local variable declaration statement, {\tool} suggests
a placeholder and consider it matching with the actual name, because
any new name can be used at that point.


To compare {\tool} with PCC~\cite{ase_17} in statement completion (SC)
on Large Corpus, we use the following SC setting that works for PCC,
which is aimed to suggest the next statement only
(Section~\ref{related}).
%
%
%
At a position $i$, the previous code sequence is divided into $C_1$=$c_1c_2...c_{t}$ and $C_2$=$c_{t+1}c_{t+2}...c_{i-1}$, where $t$ is the
ending position of the nearest completed statement. $C_1$ is used as
the input of PCC to suggest the next statement. 
We collected into the list of the resulting suggestions the top $k$
most likely code statements from PCC that begin with $C_2$. Among the
list, if there exists a statement that is the actual code sequence, we
count this as a \textit{hit}.

To compare {\tool} with PCC~\cite{ase_17} in next-statement (NS)
suggestion on Small Corpus, we uses the same setting as in
PCC~\cite{ase_17}. That is, instead of traversing source code
sequentially token by token, we ran {\tool} and computed the top-$k$
accuracy only at the beginning position of every code
statement. However, in Small Corpus, the test data includes individual
files without containing the corresponding projects' files. Meanwhile,
{\tool} is designed using program analysis on the code of currently
developing projects. Therefore, we created dummy projects for each of
the testing files.

%% file: results.tex
\section{Empirical Results}
\label{results}

\input{RQ1}

\input{RQ2}

\input{RQ3}

\vspace{-0.06in}
\subsection{Time Complexity (RQ4)}

All experiments were run on a Windows with 16 Intel Xeon
3.7GHz, 32GB RAM. {\tool} took 10 minutes for training. The
average running time for a request is 5.5s. On
average, in the results in which the remaining code sequence is in
top 5 of the ranked list, the average number of tokens in the
remaining code sequences is 3.1 tokens. This equals the typing speed
of about 0.56 tokens per sec, which is slightly slower than the
average human typing speed, 0.68 tokens/sec~\cite{typing_speed}.

\input{incorrect_cases}

\input{threats}

%% file: RQ1.tex
\subsection{Accuracy Comparison (RQ1)}

\input{RQ1_sc}


%% file: RQ1_sc.tex
\subsubsection{Comparative Results of SC on Large Corpus}
\label{compare}

We compared {\tool} with PCC~\cite{ase_17}, which applies a
statement-level $n$-gram LM and searches for similar statements for
next statement completion (will be detailed in
Section~\ref{related}). We also compared {\tool} against two baseline
approaches: $n$-gram LM and $n$-gram+PA. For $n$-gram LM, we trained a
$n$-gram LM and used it to predict the next token by token, and rank
the candidates for the code sequence
according their occurrence likelihoods. The $n$-gram+PA model works
similarly to $n$-gram LM except that
PA is additionally applied to filter out the invalid candidate
sequences. Then, the valid ones is ranked.
We used the $6$-grams for both $n$-gram LM and $n$-gram+PA. In
{\tool}'s $n$-gram LMs, $\phi_{excode}$ and $\phi_{lexemes}$, $n=6$.
We did not compare with a model that solely uses
PA since it generates a huge number of equally-ranked candidates.

\input{RQ1_SC_cross}

As seen in Table~\ref{RQ1_SC_cross}, the top-1 accuracy for {\tool} is
39.8--41.3\%. That is, up to 4 out of 10 requests, users
could find their expected next code sequence for the current statement
at the top of our ranked list.
For PCC, the top-1 accuracy is from 0.6--4.3\%, that
is {\bf 9X--69X lower than {\tool}'s}. Meanwhile,
lexical $n$-gram achieves only from 0.02--0.28\%. Even when we used PA
to filter out invalid suggestions, the top-1 accuracy is still very
low, 0.02--0.34\%, that is more than 100X lower than
{\tool}'s top-1 accuracy.
%
For top-5 accuracy, {\tool} also achieves up to {\bf 51.0\%}, which is
7--14X and more than 50X higher than PCC and both
$n$-gram LM and $n$-gram+PA.

There are two key reasons for their low accuracy. First, code statement as
its entirety is relatively project-specific~\cite{ase_17}. Indeed, on
average, the portion of the code statements in a project that can be
found in others is only 3.2\%.
That leads to the low accuracy of PCC~\cite{ase_17}, which relies on
the repetition of entire code statements.
%
%
Second, for $n$-gram baselines, because the next sequence is suggested
by predicting next token one at a time, the accuracy of next sequence
suggestion is affected by the confounding effect of the accuracy of a
single next-token suggestion.
The highest top-1 accuracy of an $n$-gram LM for next code token
suggestion is about 0.5~\cite{saner18}. Therefore, for predicting a
next code sequence containing 6 tokens (on average), the maximum top-1
accuracy is $0.5^6 \approx 1.6\%$. Note that, in this
experiment, we used Large Corpus and the statement-completion (SC)
setting that are different from Small Corpus and the next-statement
(NS) setting used in PCC~\cite{ase_17}. Thus, this leads to a
different accuracy for PCC than the one reported in its
paper~\cite{ase_17}.

\subsubsection{Comparative Results of next-statement (NS) suggestion on Small Corpus}
As seen in Table~\ref{statement_suggestion}, {\tool} does not perform
next-statement suggestion as good as PCC.
The main reason is that the test set contains individual Java files
without the project-specific files and information such as the fields and
methods of the classes. Thus, the components of {\tool} relevant
to program analysis, \eg identifying the valid candidates for the
next token, and type-checking, cannot be performed as expected.



\begin{table}[]
\centering
\caption{Next-Statement Suggestion Accuracy}
\label{statement_suggestion}
\begin{tabular}{l|l|l|l|l}
\toprule
       & Top 1  & Top 3  & Top 6  & Top 10 \\ \hline
{\tool} & 20.3\% & 28.5\% & 32.0\% & 42.2\% \\ 
PCC~\cite{ase_17}    & 28.9\% & 51.1\% & 54.8\% & 59.3\% \\ \toprule
\end{tabular}
\end{table}

\input{further_study}

%% file: RQ1_SC_cross.tex
\begin{table}[t] 
\centering
\caption{Code Statement Completion Accuracy}
\label{RQ1_SC_cross}
\begin{tabular}{ll|llll}
\toprule
Project   	 & Top-$k$   & {\tool}     
			 & \begin{tabular}[c]{@{}l@{}}$n$-gram \\ LM\end{tabular} 
			 & \begin{tabular}[c]{@{}l@{}}$n$-gram\\ LM + PA\end{tabular} 
			 & PCC~\cite{ase_17} \\ \hline

Ant       	 & \begin{tabular}[c]{@{}l@{}}Top 1\\ Top 5\end{tabular} 
			 & \begin{tabular}[c]{@{}l@{}}41.3\%\\ 48.7\%\end{tabular} 	
			 & \begin{tabular}[c]{@{}l@{}}0.09\%\\ 0.33\%\end{tabular}		
			 & \begin{tabular}[c]{@{}l@{}}0.12\%\\ 0.44\%\end{tabular}                                                                                                                       
			 & \begin{tabular}[c]{@{}l@{}}2.13\%\\ 4.65\%\end{tabular}		\\  \hline

Batik     	 & \begin{tabular}[c]{@{}l@{}}Top 1\\ Top 5\end{tabular} 
			 & \begin{tabular}[c]{@{}l@{}}39.8\%\\ 50.1\%\end{tabular}
			 &\begin{tabular}[c]{@{}l@{}}0.28\%\\ 0.78\%\end{tabular}
			 & \begin{tabular}[c]{@{}l@{}}0.34\%\\ 0.83\%\end{tabular}                                                                                                                       
			 & \begin{tabular}[c]{@{}l@{}} 4.24\%\\ 7.24\%\end{tabular}		\\  \hline

Cassandra & \begin{tabular}[c]{@{}l@{}}Top 1\\ Top 5\end{tabular} 
			 & \begin{tabular}[c]{@{}l@{}}40.5\%\\ 49.3\%\end{tabular}                                                                                                                
			 &\begin{tabular}[c]{@{}l@{}}0.19\%\\ 0.44\%\end{tabular}
			 & \begin{tabular}[c]{@{}l@{}}0.25\%\\ 0.50\%\end{tabular}                                                                                                                       
			 & \begin{tabular}[c]{@{}l@{}} 2.59\%\\ 5.89\%\end{tabular}		\\  \hline

Log4J     	 & \begin{tabular}[c]{@{}l@{}}Top 1\\ Top 5\end{tabular} 
			 & \begin{tabular}[c]{@{}l@{}}40.4\%\\ 48.2\%\end{tabular}                                                                                                                                                                       			 &\begin{tabular}[c]{@{}l@{}}0.02\%\\ 0.10\%\end{tabular}
			 & \begin{tabular}[c]{@{}l@{}}0.02\%\\ 0.15\%\end{tabular}                                                                                                                       
			 & \begin{tabular}[c]{@{}l@{}} 1.12\%\\ 3.94\%\end{tabular}		\\  \hline

Lucene    	 & \begin{tabular}[c]{@{}l@{}}Top 1\\ Top 5\end{tabular} 
			 & \begin{tabular}[c]{@{}l@{}}39.3\%\\ 49.5\%\end{tabular}                                                        
			 & \begin{tabular}[c]{@{}l@{}}0.23\%\\ 0.54\%\end{tabular}
			 & \begin{tabular}[c]{@{}l@{}}0.31\%\\ 0.61\%\end{tabular}                                                                                                                       
			 & \begin{tabular}[c]{@{}l@{}}3.87\%\\ 6.60\%\end{tabular}		\\  \hline

Maven-2   	 & \begin{tabular}[c]{@{}l@{}}Top 1\\ Top 5\end{tabular} 
			 & \begin{tabular}[c]{@{}l@{}}40.2\%\\ 51.0\%\end{tabular}                                                        
			 & \begin{tabular}[c]{@{}l@{}}0.10\%\\ 0.25\%\end{tabular}
			 & \begin{tabular}[c]{@{}l@{}}0.16\%\\ 0.26\%\end{tabular}                                                                                                                       
			 & \begin{tabular}[c]{@{}l@{}}0.58\%\\ 3.32\%\end{tabular}		\\  \hline

Maven-3   	 & \begin{tabular}[c]{@{}l@{}}Top 1\\ Top 5\end{tabular} 
			 & \begin{tabular}[c]{@{}l@{}}38.9\%\\ 48.7\%\end{tabular}                                                        
			 &\begin{tabular}[c]{@{}l@{}}0.16\%\\ 0.28\%\end{tabular}
			 & \begin{tabular}[c]{@{}l@{}}0.22\%\\ 0.38\%\end{tabular}                                                                                                                       
			 & \begin{tabular}[c]{@{}l@{}}1.33\%\\ 4.86\%\end{tabular}		\\  \hline

Xalan     	 & \begin{tabular}[c]{@{}l@{}}Top 1\\ Top 5\end{tabular} 
			 & \begin{tabular}[c]{@{}l@{}}40.1\%\\ 49.2\%\end{tabular}                                                        
			 &\begin{tabular}[c]{@{}l@{}}0.17\%\\ 0.41\%\end{tabular}
			  & \begin{tabular}[c]{@{}l@{}}0.19\%\\ 0.63\%\end{tabular}                                                                                                                       
			 & \begin{tabular}[c]{@{}l@{}}2.91\%\\ 6.12\%\end{tabular}		\\  \hline

Xerces    	 & \begin{tabular}[c]{@{}l@{}}Top 1\\ Top 5\end{tabular} 
			 & \begin{tabular}[c]{@{}l@{}} 39.7\%\\ 49.9\%\end{tabular}                                                        
			 &\begin{tabular}[c]{@{}l@{}}0.22\%\\ 0.49\%\end{tabular}
			 & \begin{tabular}[c]{@{}l@{}}0.23\%\\ 0.61\%\end{tabular}                                                                                                                       
			 & \begin{tabular}[c]{@{}l@{}} 2.35\%\\ 6.31\%\end{tabular}		\\ 
			 \toprule

\end{tabular}
\end{table}

%% file: further_study.tex
\subsubsection{Analysis}


We analyzed the correct results and found that {\tool}'s high accuracy
can be attributed to the following. {\em First, it uses {\em excode}
to derive the template at a higher abstraction level}.  This helps
{\tool} learn code patterns from other locations and avoid missing the
potential candidates for the next sequence for the given partial code.
%
%
For example, given a partial statement which starts
with \code{cfnames.add(} at line 4 of Figure~\ref{example3},
where \code{cfNames} is a \code{List$<$String$>$}, its expected next
sequence is \code{cfStore.getColumnFamilyName());},
where \code{cfStore} is a local variable. The partial statement and
the expected sequence both have never appeared in the training
data. This makes the baseline models, which work on lexical tokens,
fail. Meanwhile, the {\em excode} sequence corresponding
to \code{cfnames.add(} and that {\em excode} sequence
for \code{cfStore.getColumnFamilyName()} co-occur several
times. Thus, \code{VAR(ColumnFamilyStore) OP(ACC)
CALL(getColumnFamilyName)...LP RP RP ;} (corresponding to the code
sequence \code{cfStore.getColumnFamilyName());}) is listed in the set
of about 2,000 candidate templates.
%

Additionally, {\em the application of PA to filter out the
type-incorrect templates help {\tool} achieve high accuracy}. In the
above example, the set of 137 valid candidates among 2,000 candidates,
that are learned from other places, are adapted to fit with the
current context using PA. For example, {\tool}
concretized \code{VAR(ColumnFamilyStore) OP(ACC)...LP RP RP ;} by
using the accessible variable \code{cfStore} for \textit{excode}
token \code{VAR(ColumnFamilyStore)} instead of \code{cfs}
or \code{filter} as in other models. \textit{This adaptation ability
to the current method with PA is the third reason for {\tool}'s high
accuracy}.

\begin{figure}[t]
\begin{center}
\begin{lstlisting}[captionpos=b, basicstyle={\footnotesize\ttfamily},numbersep=-4pt]
List<String> cfNames = new ArrayList<String>();
Set stores = getValidColumnFamilies(columnFamilies);
for (ColumnFamilyStore cfStore : stores) {
	cfNames.add(	
	//cfStore.getColumnFamilyName());
		
\end{lstlisting}
\vspace{-0.15in}
\caption{A partial method in Cassandra}
\label{example3}
\end{center}
\end{figure}

Another reason for our high accuracy is that \textit{in {\tool},  OOV is addressed by enforcing accessibility rules to avoid missing the
valid file-specific or project-specific tokens when producing the
candidate templates.}

Finally, {\tool} leverages the naturalness of source code
in the lexical form,
to effectively rank the candidate code sequences.
In Figure~\ref{example2}, given a partial statement starting with 
\code{reports.addAll(}, where the type of \code{reports} is \code{List},  
the expected sequence is \code{getReportExecutions());}. In fact, method 
\code{getReportExecutions} is declared inside the current class and has 
never been seen in the training data. This accessible call is still used to 
produce template. Since the type restriction for the argument of method 
\code{addAll}, there are a few type-valid candidates, such as 
\code{getReportExecutions());} or \code{null);}. Then, the candidate 
\code{getReportExecutions());} is ranked on the top by the lexeme-based LM, 
$\phi_{leximes}$, because the tokens \code{reports}
and \code{Report} in \code{getReportExecutions} frequently go
together, such as: \code{\textit{reports}.contains(\textit{report})}
and \code{\textit{reports}.add(\textit{report}Mojo);}.

\begin{figure}[t]
\begin{center}
\begin{lstlisting}[captionpos=b, basicstyle={\footnotesize\ttfamily},numbersep=-4pt]
List<MojoExecution> reports = new ArrayList();
reports.addAll( //getReportExecutions();
\end{lstlisting}
\vspace{-0.15in}
\caption{A partial method in Apache Maven}
\label{example2}
\end{center}
\end{figure}

\input{problem_start_stm}

%% file: problem_start_stm.tex
We further studied the cases in which {\tool} did not suggest well. We
found that the majority of them are the cases whose the completion
position is near the beginning of the current statement, especially
the cases of suggesting entire statement (will be explained in
Section~\ref{position}).
Since the next-token prediction accuracy is not 100\%, the more next
tokens predicted, the lower next-sequence completion accuracy.

%

%% file: RQ2.tex
\subsection{Intrinsic Evaluation Results (RQ2)}
\label{RQ2}


We further studied the {\em complexity} and {\em diversity}, and
{\tool}'s {\em effectiveness} on different {\em kinds of code tokens}
and different {\em lengths} of the statements completed by {\tool}. We
randomly sampled 10,000 results from 460K total results.

\begin{figure}
\centering
\includegraphics[width=3.2in]{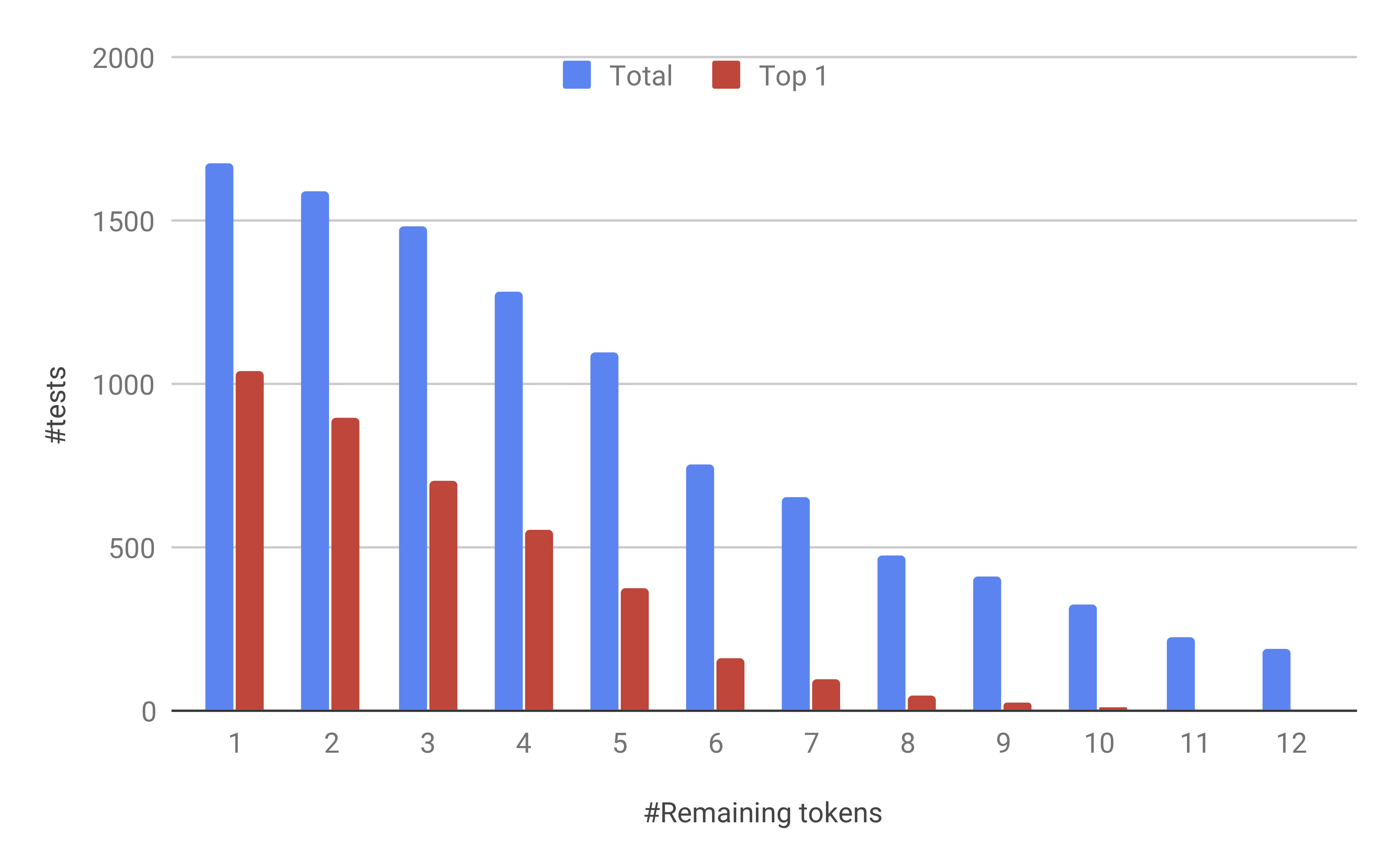} 
\vspace{-0.05in}
\caption{Accuracy on length of remaining code sequences}
\label{intrinsic_len}
\end{figure}

First, we classified the sampled results into 12 categories
corresponding to the size (1--12 tokens) of the remaining code
sequence of the currently completed statement (the maximum number of
tokens to be completed is set to 12). Figure~\ref{intrinsic_len} shows
the number of correct results over the total number of results for
each category. As expected, the longer the remaining sequence, the
more number of tokens to be completed, the less number of correct
results. As seen, {\tool} correctly handles complex completed
statements with various lengths. Also, through the similar shapes of
two types of columns from left to right, we see that the proportions
of correct results over the total ones for all categories are quite
uniform. Thus, our model \textit{is effective for various lengths of
  the remaining sequences, even for long sequences}. A correct example
in \code{Maven-3} is as follows. The given partial code is the fragment
\code{Activ} \code{activ} = \code{new Activ();$\_$}. The correct
suggestion is
\code{activ.setActiveByDefault(settingsActivation.isActiveByDefault());},
which has a total of 11 tokens after the cursor.





\begin{figure}
\centering
\includegraphics[width=3.2in]{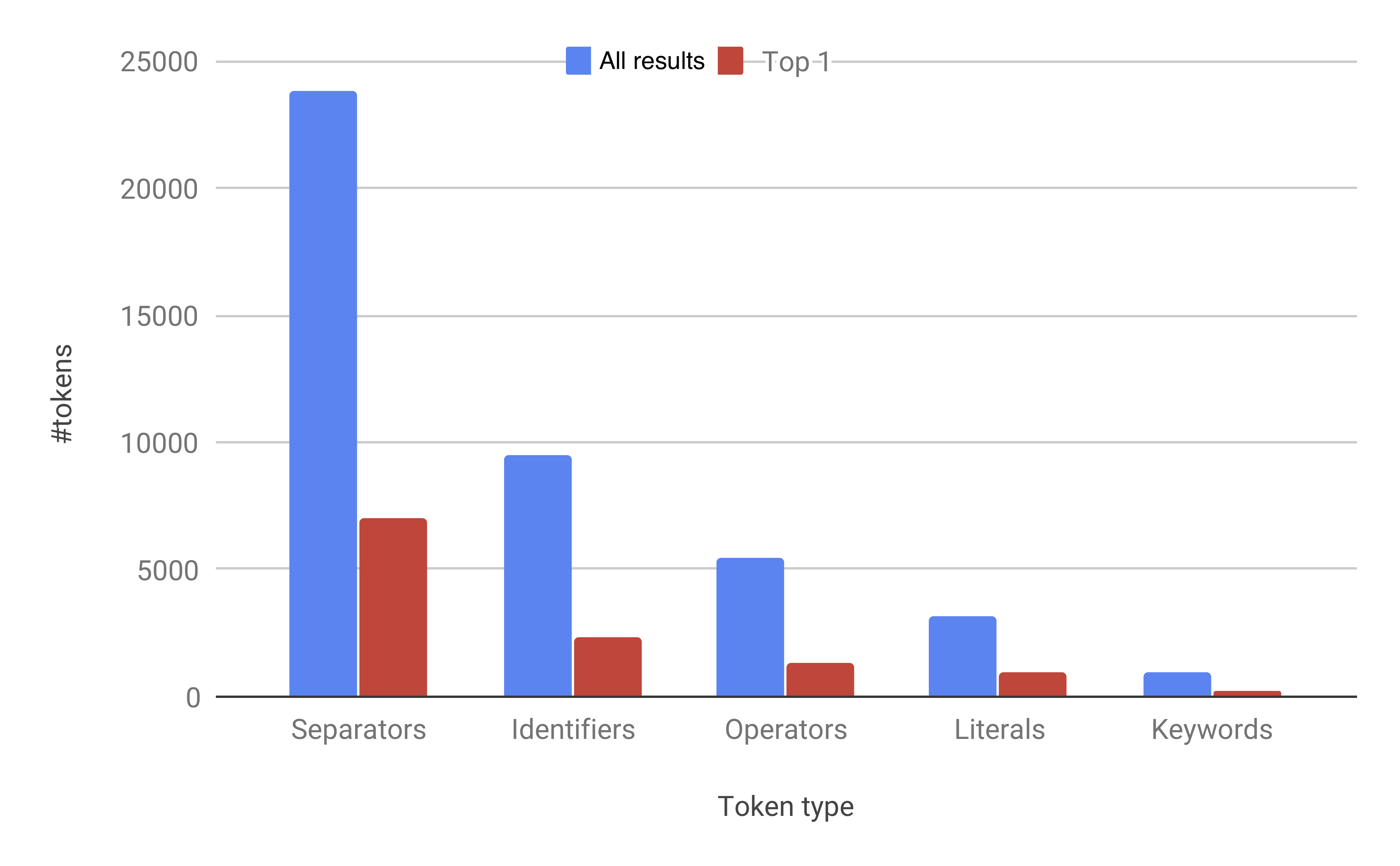}
\vspace{-0.05in}
\caption{Accuracy on various token types}
\label{intrinsic_token_type}
\end{figure}

Second, to study the results by {\tool} with respect to different
kinds of tokens, we classified all the tokens in the sampled results
into several categories corresponding to different syntactical token
types. As seen in Figure~\ref{intrinsic_token_type}, the proportions
of the correct results over the total ones for all categories are
relatively similar. Thus, {\em {\tool} is equally effective for
  diverse kinds of tokens}. A correct example is a conditional
expression, \code{null \&\& file.isDirectory())} (containing
a \code{null} literal, infix operators, identifiers, and separators.)
for a partial condition of an \textit{if-then} statement, \code{if
  (jarFile ==$\_$}.





%
%

%% file: RQ3.tex
\subsection{Sensitivity Results (RQ3)}

\subsubsection{Completion position}
\label{position}

Because {\tool} is based on the given code sequence, a
completion point in the code sequence of a method $M_n=c_1c_2...c_n$
has impact on accuracy. Thus, we conducted an experiment
to measure that. We first chose a random project, \code{Lucence}. For
each method, we chose a completion point at three locations: the first
quartile point $l_1=\floor*{n/4}+1$, the middle point
$l_2=\floor*{n/2}+1$, and the third quartile point
$l_3=\floor*{3n/4}+1$.


\input{location_impact}

As seen in Table~\ref{location_impact}, accuracy slightly
increases if we move the point to a later part of a method from
$1^{st}$ to $3^{rd}$ quartile point. This is expected as {\tool} has more information.

\begin{figure}[t]
\centering
\includegraphics[width=3.2in]{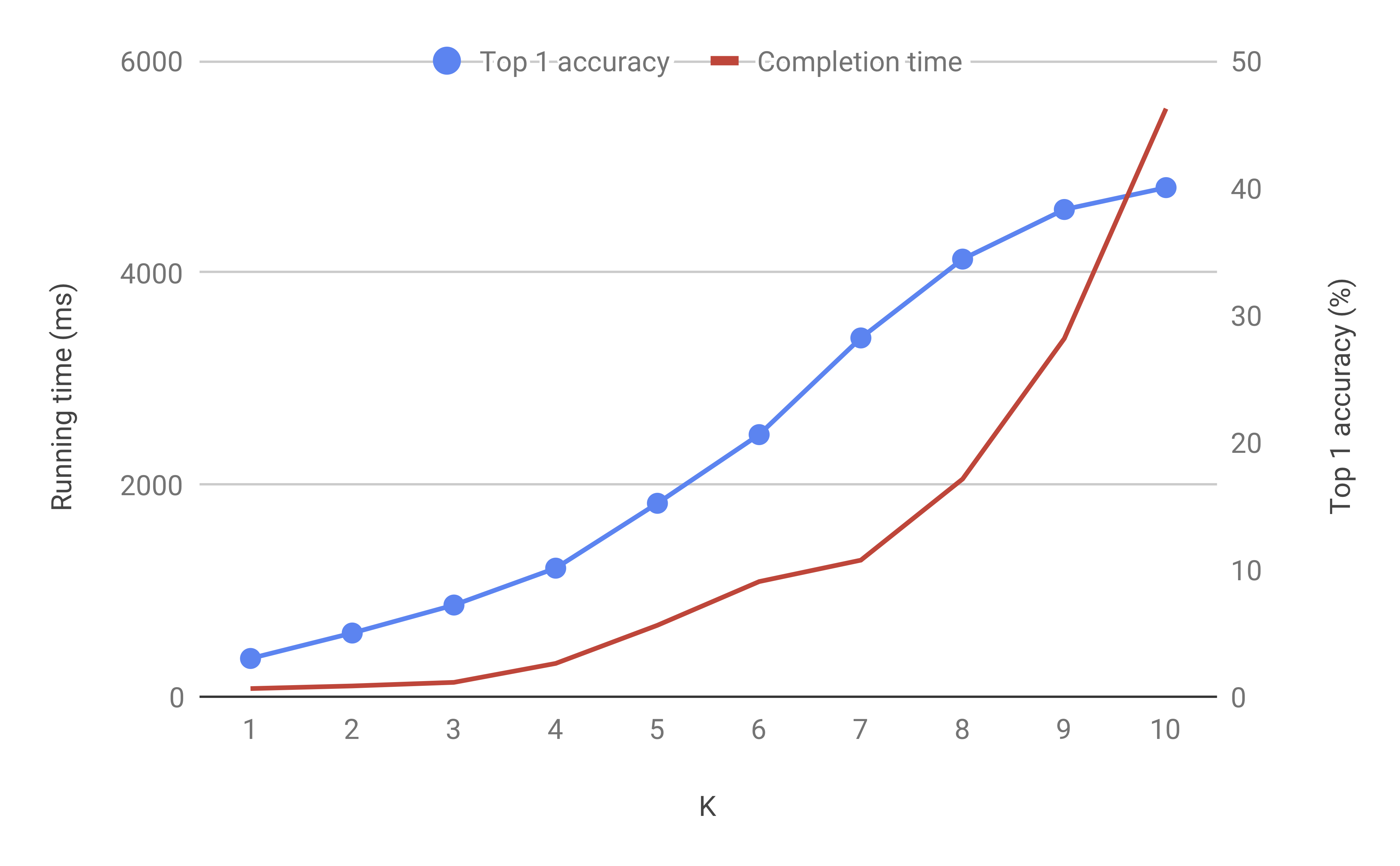}
\vspace{-0.05in}
\caption{Impact of threshold $K$ on accuracy and running time}
\label{k_time_accuracy}
\end{figure}

\begin{table}[t]
\centering
\caption{Impact of $n$ in $n$-gram LMs on Accuracy}
\label{n_gram}
\begin{tabular}{l|lllllllll}
\toprule
$n$ 	& 2		 & 3		  & 4  	   & 5  		& 6 			\\ \hline
Top 1 		& 29.7\% & 30.3\% & 31.3\%  & 35.4\% & 40.4\% 	\\
Running time 	& 1095ms & 2287ms & 3169ms  & 4388ms & 5447ms  	\\ \toprule
\end{tabular}
\end{table}

\begin{table}[t]
\centering
\caption{Impact of Training Data's Size on Accuracy}
\label{data_size}
\begin{tabular}{l|lllllllll}
\toprule
\#Folds 	& 1		 & 3		  & 5  	   & 7  		& 9 			\\ \hline
Top 1 	& 28.6\% & 31.4\% & 34.9\%  & 37.8\% &  41.3\% 	\\
Top 5 	& 30.2\% & 35.4\% & 38.5\%  & 42.0\% &  48.7\%  	\\ \toprule
\end{tabular}
\end{table}


We also computed the accuracy as the completion points at the
beginning of a new statement (\ie next-statement suggestion). The
percentage of the cases in~which the next statement being correctly
ranked on the top of the suggestion list is 8.2\% (top-1 accuracy, not
shown). This is expected because any type of statement is valid at
those beginning points. However, {\tool}'s accuracy is still 3.9X
better than top-1 accuracy in next-statement suggestion of
PCC~\cite{ase_17}.


\subsubsection{Threshold $K$}
{\tool} also relies on the pre-defined number $K$ of most likely
tokens for next \textit{excode} to identify templates.
Figure~\ref{k_time_accuracy} shows the accuracy and running time per
completion request when we varied $K$. As seen, when $K$ is small, the
accuracy are very low because the correct next \textit{excode} token
might be dropped out of top $K$. The accuracy increases when we keep a
larger number of top candidates. However, with a large $K$, $K$=9--10,
the number of the predicted code sequences is very large. This lead to
a slower increasing trend as $K$ is larger. Regarding the running
time, since the number of the predicted code sequences exponentially
increases when we increase $K$, the running time for each 
request also grows exponentially when $K$ is larger.

\subsubsection{Value of $n$ in the $n$-gram LMs}

We also measured the impact of the size $n$ in the $n$-gram LMs
$\phi_{excode}$ and $\phi_{lexemes}$, on {\tool}'s accuracy. We varied
$n$ for both $\phi_{excode}$ and $\phi_{lexemes}$ from 2--6 and
computed top-1 accuracy when we ran {\tool} on a randomly selected
project.


As seen in Figure~\ref{n_gram}, the accuracy grows from 29.7\% to
40.4\% for $n$=2--6. The reason is that the $n$-gram LMs with larger
$n$ is able to capture more precisely the current context and rank
better the correct next \textit{excode} tokens (for $\phi_{excode}$)
and the correct next code sequences (for $\phi_{lexemes}$). Meanwhile,
the running time for each completion request increases linearly from
1,095ms to 5,447ms, because longer sequences need to be computed as $n$
is increased from 2 to 6.

\subsubsection{Training data's size}

For training data's size, we randomly selected a
project, \code{Ant}, and divided its source files into 10 folds. We
used one fold for testing and increased the sizes of the
training data by adding into a dataset of 8 other projects one fold
at a time until 9 remaining folds are added.
Top-1 accuracy increases from 28.6\% to 41.3\% when we increase 
training data (Table~\ref{data_size}). As expected, with
larger training data sets, the model has observed more and
performs better.


%% file: location_impact.tex
\begin{table}[t]
\centering
\caption{Impact of Completion Points on Accuracy}
\label{location_impact}
\begin{tabular}{l|lllll}
\toprule
Location     	& Top 1 		& Top 2 		& Top 3 		& Top 4 		& Top 5 \\ \hline
1st quartile 	& 24.5\%    	&  30.1\%   	& 35.8\%   	& 39.2\% 	&  41.1\%      \\ 
2nd point 		& 38.6\%    	&  43.0\%  	& 46.6\%   	& 48.9\% 	&  51.2\%     \\ 
3rd quartile 	& 56.8\%    	&  58.6\%   	& 59.0\%    	& 60.6\% 	&  62.4\%     \\ \toprule
\end{tabular}
\end{table}


%% file: incorrect_cases.tex
\subsection{Ineffective Cases}

For incorrect cases, we classified them into the categories based on
their number of code tokens that are in the remaining of the expected
sequences.
We found that the portion of the cases which contain redundant tokens is
up to 23\%. For the percentages of the cases of 1, 2 and 3
missed-tokens are 29\%, 12\% and 26\%, respectively. Meanwhile, the
portion of the cases of more than 3 missed-tokens is only 10\%.
For example, the correct one is \code{commits.get(readFrom);} (the
suggested one is \code{commits.get(writeTo);}) for the given partial
code \code{commits =}$\_$.
\textit{Thus, these results show that even for the ineffective cases,
  {\tool}'s suggestion lists are still reasonable}.

%% file: threats.tex

\vspace{-0.06in}
\subsection{Threats to Validity}

Our selected projects are not representative and different from
PCC~\cite{ase_17}'s dataset. However, we chose a high number of
projects with large numbers of statements. For PCC, we used its
default setting for the comparison. Our simulated code suggestion
procedure is not true code editing. Inaccuracy is from the fact
that {\tool} cannot correctly resolve types/roles sometimes due to
incomplete code.

%% file: related.tex
\section{Related Work}
\label{related}


{\tool} is related to PCC by Yang {\em et al.}~\cite{ase_17}. In
comparison, there are fundamental differences between {\tool} and
PCC. First, PCC focuses on suggesting the next statement when a user
finishes the previous statement, while {\tool} supports both filling a
partially typed statement (SC) and generating a next statement
(NS). PCC can be used to support statement completion when the
partially typed statement is matched against the suggested statement
$s$, and the remaining tokens of $s$ will be recommended for users.
Second, while PCC is based solely on statistical LM, {\tool} combines
PA and LM.
Third, the way PCC used an LM is also different. PCC combines all
lexical tokens belonging to a statement into a pseudo-token called
{\em IR}, for the statement. In training, it converts source code into
sequences of IRs and trains a $n$-gram model to learn to recommend an
entire statement. Because the entire statements do not repeat often,
PCC has to consider similar IRs as the same, causing
inaccuracy. {\tool} uses LM+PA to predict token by token and compose
them. We showed that {\tool} outperforms PCC in both SC and NS.

There exists a rich literature of approaches on CC. The
approaches can be broadly classified into the following categories.
The first category relies on program analysis.
IDEs support the completion of method calls/field accesses.
%
Eclipse~\cite{eclipse} and IntelliJ IDEA~\cite{intellisense,informer}
also support {\em template-based} completion for common constructs and
APIs (\code{for}/\code{while}, \code{Iterator}).

The second category uses code
pattern mining~\cite{bruch2009,hill04,strathcona05,hou-icsm11,mylyn06,icse12-grapacc,omar12,robbes08,zhang-icse12,zhong2009}. Grapacc~\cite{icse12-grapacc}
uses API patterns to match them against the current
code.
%
Bruch {\em et al.}~\cite{bruch2009} suggest a call
based on frequent methods, co-occurrent calls, and best
matching and their calling structures.

The third category relies on statistical LMs~\cite{manning99}.  Hindle {\em et al.}~\cite{prem_naturalness}
use $n$-gram on lexical tokens to predict the next token.
Later, Tu {\em et al.}~\cite{tu-fse14} improve $n$-gram model with
caching for recently seen tokens. Raychev {\em et
al.}~\cite{ethz-pldi14} use $n$-gram to
predict API call. SLAMC~\cite{fse13} associates code tokens with {\em
sememes}, including token roles and data types. In comparison, there
are key differences.
First, {\em excode} is designed for template statements while sememes
are abstractions over source code to predict the next token. Second,
{\tool} has a type checker for {\em excode} with \code{Unknown}
type, while {\em sememes} do not have it. Third, $n$-gram topic model
is used in {\em sememes} to provide the context for prediction, while
{\tool} uses PA+LM. Finally, SLAMC suggests only the next token.
GraLan~\cite{icse15} is a graph-based LM that captures usage
patterns to suggest API calls.

Recent advances in deep learning have been used in next token suggestion.
White {\em et al.}~\cite{white-msr15} use
Recurrent Neural Network (RNN) to learn the context to predict the
next token, while Dam {\em et al.}~\cite{dam16} rely on LSTM.
DNN4C incorporates syntactic information for
better prediction using DNN LM~\cite{saner18}.

Despite the success of using statistical LMs, those
existing approaches are still limited to support only next token.
They do not combine LM with PA as in {\tool}.

%% file: conclusion.tex
\section{Conclusion}
We introduce {\tool}~\cite{website1}, which combines PA
and the principle of software naturalness complete partial statements.
We aim to benefit from the strengths of both directions.
%
%
{\tool} is trained on a code corpus to learn the candidate
templates. Then, it uses PA to validate and concretize the templates
into valid code statements. Finally, they are ranked by using a LM
trained on the lexical form of the source code.
%

We conducted several experiments to evaluate {\tool} in statement
completion and next-statement suggestion on datasets with +460K
statements with a total of +1M suggestion points. Our results show
that {\tool} is very effective with top-1 accuracy of 40\% and top-5
accuracy of 49.4\% on average. That is, in 4 out of 10 cases, when a
user requests to complete his/her currently-written statement, (s)he
can find the remaining of the desired statement in the top of the
suggestion list.
Importantly, {\tool} significantly improves over the baseline model
using only $n$-gram on lexical code (up to 142X in top-1
accuracy) and the model using lexical $n$-gram+PA (up to 117X in
top-1 accuracy). It also improves over the state-of-the-art tool
PCC~\cite{ase_17} with 69X higher in top-1 accuracy.